\input colordvi

\documentstyle[jkas]{article}

\beginpage{1}
\endpage{9}
\year{2013}\volume{00}\month{Dec}
\runningauthor {S.-S. Lee} 
\runningtitle{Intrinsic Brightness Temperature}

\month{Dec} \year{2013} \volume{46} \issueno{6}
\beginpage{1}\endpage{9}
\date{Received 2013 September 13; Revised 2013 October 29; Accepted 2013 November 4}

\begin{document}
\title{INTRINSIC BRIGHTNESS TEMPERATURE OF COMPACT RADIO SOURCES AT 86GHZ}
\author{Sang-Sung Lee} 
\address{$^1$ Korea Astronomy and Space Science Institute,
776 Daedukdaero (61-1 Hwaam), Yusong, Daejon 305-348, Korea\\
 {\it E-mail : sslee@kasi.re.kr}}

\address{\normalsize{\it (Received 2013 September 13; Revised 2013 October 29; Accepted 2013 November 4)}}

%--------------------------------------------------------------------
\abstract{We present results on the intrinsic brightness temperature
of a sample of compact radio sources observed
at 86~GHz using the Global Millimeter VLBI Array. 
We use the observed brightness temperatures at 86~GHz
and the observed superluminal motions at 15~GHz for the sample
in order to constrain the characteristic intrinsic brightness temperature
of the sample.
With a statistical method for studying the intrinsic brightness temperatures
of innermost jet cores of compact radio sources,
assuming that all sources have the same intrinsic brightness temperature
and the viewing angles of their jets are around the critical value
for the maximal apparent speed,
we find that sources in the sample have a characteristic
intrinsic brightness temperature,
$T_{\rm 0} = 4.8^{+2.6}_{-1.5}\times 10^{9}$\,K,
which is lower than the equipartition temperature for the condition that 
the particle energy equals to the magnetic field energy. 
Our results suggest that the VLBI cores seen at 86~GHz
may be representing a jet region where the magnetic field energy
dominates the total energy in the jet.
}

\maketitle

%--------------------------------------------------------------------
\section{INTRODUCTION}

Compact radio sources are generally defined as
radio sources whose flux density at
an intermediate radio frequency, e.g., $\sim1$~GHz, is
dominated by
the emission of a single bright region within $\sim1$~kpc in size~\citep{BK79}. 
Compact radio sources
usually have flat radio spectra and exhibit pronounced radio variability.
Moreover, due to the lager ratio of optical to radio flux
of the compact radio sources than of steep-spectrum sources, 
these objects have been easily identified. 

Radio emission from parsec-scale jets in compact radio sources 
consists of optically thin synchrotron emission,
and characterized by their spectral and polarization properties
and significant inverse-Compton emission~\citep[see][]{mar90,HM91}.
The flat spectrum of the radio emission is generally interpreted 
as due to superposition of incoherent
synchrotron radiation from a non-thermal distribution of
relativistic electrons located in several distinct
components~\citep{KP69,mar95}. 
These components form
the innermost compact structure,
the compact jet base at sub-parsec scales,
and the bright emission regions at parsec scales of the jet.
The physical processes of the formation of inner jets that connect 
the nucleus to the observed radio jet, their acceleration 
to relativistic speeds, 
and strong collimation to large scales (pc to kpc)
have been extensively investigated 
but remain poorly understood~\citep[e.g.,][]{mar06,LZ06,lob07}. 

\citet{rea94} suggested that parsec-scale jets are in an equipartition
condition that the energy density in relativistic particles is equal
to that in magnetic fields~\citep{bb57}.
He found that the radio sources may radiate at an equipartition
brightness temperature around $5\times10^{10}$~K in most circumstances.
Recent studies have shown that extended radio lobes
are indeed at equipartition~\citep{cro+05},
and the relativistic jets are at equipartition
in their median-low state~\citep{hom+06}.
However when the relativistic jets are in their maximum state,
the brightness temperatures are a factor of 4 larger,
implying the energy in their radiating particle is $\geq10^{5}$ times larger
than the energy in magnetic fields, based on the relation of  
the energy and brightness temperature as proposed
in \citet{rea94}: $\eta = u_{p}/u_{B} =(T_{\rm eq}/T_{\rm b})^{-17/2}$,
where $u_p$,$u_B$ are the energy densities of the radiating particles
and the magnetic field, respectively, $T_{\rm eq}\simeq 5\times 10^{10}$~K
is the equipartition brightness temperature,
and $T_{\rm b}$ is the observed brightness temperature.

It is very difficult to measure intrinsic properties of
extragalactic compact radio
sources because the jets of compact radio sources are highly relativistic 
and therefore Doppler boosted~\citep{BK79,LB85}.
The physical aspects of the jet can be 
parameterized by the Lorentz factor $\gamma_{\rm j}$, the intrinsic
brightness temperature $T_{\rm 0}$, and the angle to the line of sight
$\theta_{\rm j}$. From these intrinsic physical properties, one can calculate
the Doppler factor $\delta$, the apparent jet speed $\beta_{\rm app}$, and 
the observed brightness temperature $T_{\rm b}$:
   \begin{equation}
   \label{eqn:Tb-1}
   \delta = \frac{1}{\gamma_{\rm j} (1-\beta {\rm cos}\theta_{\rm j})},
   \end{equation}
   \begin{equation}
   \label{eqn:Tb-2}
   \beta_{\rm app} = \frac{\beta {\rm sin}\theta_{\rm j}}{1-\beta {\rm cos}\theta_{\rm j}},
   \end{equation}
   \begin{equation}
%  T_{\rm b} = T_{\rm 0} \delta^{n},
   T_{\rm b} = T_{\rm 0} \delta,
   \label{eqn:Tb-3}
   \end{equation}
where $\beta = (1-{\gamma_{\rm j}}^{-2})^{1/2}$ 
is the speed of jet in the rest frame
of the source (units of $c$).

Observed brightness temperatures ($T_{\rm b}$) of compact radio jets
can be used 
to study the intrinsic physical properties of the relativistic jets. 
One application is to test accelerating and decelerating
jet models~\citep{mar95}
by investigating the change in the observed brightness temperatures
measured at various radio frequencies (e.g., 2 -- 86\,GHz).
Under equipartition conditions between jet particle and magnetic field
energy densities, the position shift of the brightest jet components
of VLBI images (VLBI cores) between two frequencies
can be predicted~\citep{lob98}. The brightness temperatures in the rest
frame of the sources and the predicted core shift should be able to 
test the inner jet models.
%(Lee et al. in prep).

Another application is to obtain the intrinsic brightness temperatures of VLBI
cores by using the observed brightness temperature $T_{\rm b}$ and 
the maximum apparent jet speed $\beta_{\rm app}$.
A method developed by \cite{hom+06} was applied to the 2\,cm survey
data~\citep{kel+04} in order to determine the intrinsic physical
properties of prominent AGN~\citep{coh+07}.
They found that VLBI cores observed at 15~GHz are 
near equipartition in their median--low state, 
resulting in intrinsic brightness temperatures of 
$T_{\rm 0} = 3\times 10^{10}$\,K. 

The method can be applied to VLBI survey data at different frequencies,
with the maximum apparent jet speeds taken from~\cite{kel+04}, 
in order to constrain the intrinsic brightness temperature at these frequencies. 
In this paper, we combine the observed brightness temperature data
from a global 86~GHz VLBI imaging survey~\citep{lee+08}
and proper-motion data from the 2~cm VLBA survey.
In section 2, the theory and its application to the 2~cm VLBA survey are
reviewed. In section 3, we present our results using the 86~GHz VLBI survey.
In section 4, we discuss the interpretation of the intrinsic brightness
temperatures at various frequencies.
   
\begin{figure}[!t]
 	\centering 
	\epsfxsize=8cm
	\epsfbox{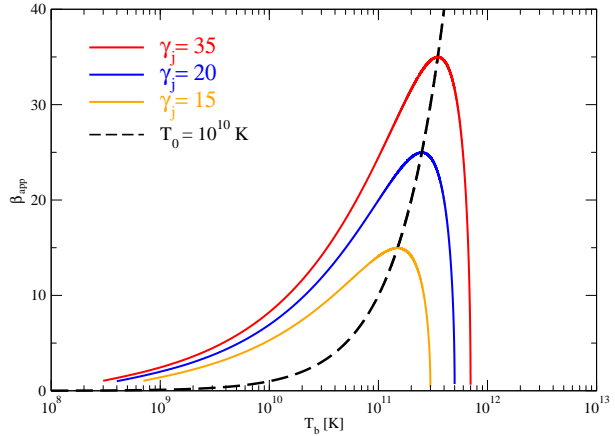}
   	\caption{ 
            Plot of the apparent jet speed $\beta_{\rm app}$
            with the observed brightness temperature $T_{\rm b}$
            for a single intrinsic brightness temperature of 
            $T_{\rm 0} = 1\times 10^{10}\,{\rm K}$ and several
            values of Lorentz factors 
            $\gamma_{\rm j} = 15, 20, 35$.
            The dashed line represents where sources observed at 
            the critical angle would lie on this plot. 
            The solid line represents the possible apparent speeds of 
            a $\gamma_{\rm j}$ source with intrinsic brightness temperature 
            given by $T_0$.
   	}
   	\label{fig1} 
\end{figure}

\section{Intrinsic brightness temperature at 15~GHz}

\subsection{Methodology}

   % From Homan et al. 2006
Following~\citet{hom+06}, we assume that 
a compact radio source contains an ideal relativistic jet, 
which is narrow and straight with 
no bends between the VLBI core and the jet components. 
Of course, some jets are not straight 
and $\theta_{\rm j}$ is not the same in the core and 
in the moving jet components.
The celebrated example and evidence of the jet bending are found
in 3C~279~\citep{hom+03,abd+10}. However, as long as superluminal motion
is observed, 
the motion must be close to the line of sight, 
and angular changes of the motion could be highly magnified due to
projection effect. 
A jet with an intrinsic bend of only a few degrees  
could be observed as a right-anlge bending jet~\citep{coh+07}. 

In this case, we can also assume that 
the maximum speed of the jet component is the same as the speed of 
the jet flow through the jet core. The flow speed of the jet is usually
different from the pattern speed of the jet in some low-luminosity sources.
However, in those sources which are bright and straight, the pattern speeds
are the same as the flow speeds.
For simplicity, we make two further assumptions: 
   \begin{enumerate}  
   \item{The intrinsic brightness temperature $T_{\rm 0}$ of all jets are the same},
and
   \item{The viewing angels of their jets are around the critical value
      $\theta_{\rm c} = {\rm arccos}\beta$ for the maximal apparent speed at 
      a given $\beta$}.
   \end{enumerate}  
Under the assumptions above, one can relate the observed brightness temperature 
to the maximum jet speed: 
   \begin{equation}
   \label{eqn:delta}
   \delta \simeq \beta_{\rm app}
   \end{equation}
and
   \begin{equation}
   \label{eqn:Tb-Bapp-To}
   T_{\rm b} \simeq \beta_{\rm app} T_{\rm 0}. 
   \end{equation}
This resultant simple relation between
the observed brightness temperature and the apparent maximum jet speed 
is illustrated in Figure~\ref{fig1}. 
From equations (\ref{eqn:Tb-1}) and (\ref{eqn:Tb-2}), one can also
relate the apparent jet speed $\beta_{\rm app}$ to the Doppler factor $\delta$
and the Lorentz factor $\gamma_{\rm j}$:
   \begin{equation}
   \label{eqn:Tb-4}
   \beta_{\rm app} = \sqrt{ (\delta \gamma_{\rm j} \beta)^2 - (\delta \gamma_{\rm j} - 1)^2 }.
   \end{equation}
This relation is illustrated as three solid lines in Figure~\ref{fig1},
for the maximum and minimum possible Doppler factors 
$\delta_{\rm max} = 1/\gamma_{\rm j}$ and 
$\delta_{\rm min} = 1/(\gamma_{\rm j} - \sqrt{{\gamma_{\rm j}}^2 -1} )$.
The lines show
actually the apparent speeds as functions of $T_{\rm b}$
for jets with Lorentz factors
of $\gamma_{\rm j} = $15, 20, and 25.
\cite{hom+06} found that from the simulation of a relativistically beamed
population of 1000 fictional compact radio sources with given $T_{\rm 0}$
and $\gamma_{\rm j}$, 
the dashed line divided the 1000 compact radio sources into two groups:
one group of about 750 sources in the right and below,
and another group of about 250 sources in the left and above.
The viewing angles of 750 sources in the right and below are smaller than
the critical angle, and their Doppler factors are large.
Their proper motions are small due to their viewing angle. 

\subsection{Application to 15~GHz data}

By combining the 2~cm VLBA survey data and the proper motion data
of AGN jets, \cite{hom+06} found
an intrinsic brightness temperature of $T_0 \simeq 3\times10^{10}$~K
of the sources in their sample when the sources are
in their median-low (25\%-median) brightness temperature state, 
that is for median brightness temperatures of a lower half sample of 
observed brightness temperaturs obtained from multi-epoch observations
of individual AGN. 
This value for $T_0$ is close to
the equipartition temperature under the condition that 
the particle energy equals to the magnetic field energy. 
However, for maximum observed brightness temperatures,
they also found
a characteristic intrinsic brightness temperature
of $2\times10^{11}$~K of their sample, which is brighter than
the equipartition temperature by a factor of 4.
This implies that, in the maximum brightness state,
the energy in radiating particles exceeds the energy in the magnetic field
by a factor of $\sim 10^5$.
They suggest that
at the innermost regions of the jet,
injection or acceleration of particles should
maintain the energy in the radiating
particles dominant over the energy in the magnetic field.

\begin{deluxetable}{cccrrrr}
\tablecolumns{7}
\tablewidth{0pc}
\tablecaption{Observed Brightness temperatures\label{tbl1}}
\tablehead{
\colhead{} & 
\colhead{} & 
\colhead{} &
\colhead{$\beta_{\rm app}$}    & 
\colhead{$T_{\rm b}^{15{\rm GHz,max}}$}    & 
\colhead{$T_{\rm b}^{15{\rm GHz,25\%med}}$}    & 
\colhead{$T_{\rm b}^{86{\rm GHz}}$}\\
\colhead{Name} & 
\colhead{Type} & 
\colhead{$z$} &
\colhead{[$c$]}    & 
\colhead{[K]}    & 
\colhead{[K]}    & 
\colhead{[K]}
}
\startdata
%here Ba-Tb.2.tableout 50 lines
0003$-$066 &B &0.347 &  8.39 $\pm$   0.39 &$>$4.60e+11 &$ $8.07e+10 &$>$2.60e+10 \\ 
0007$+$106 &G &0.089 &  1.20 $\pm$   0.07 &$ $9.05e+11 &$ $1.82e+11 &$>$1.20e+10 \\ 
0016$+$731 &Q &1.781 &  8.23 $\pm$   0.34 &$>$7.24e+12 &$ $1.14e+11 &$>$1.80e+11 \\ 
0048$-$097 &B &  ... &  0.00 $\pm$   0.00 &$>$7.31e+12 &$>$2.44e+11 &$ $2.50e+10 \\ 
0106$+$013 &Q &2.107 & 24.40 $\pm$   3.90 &$>$3.87e+12 &$ $8.83e+11 &$ $1.70e+11 \\ 
0119$+$041 &Q &0.637 &  0.50 $\pm$   1.60 &$>$1.86e+11 &$ $1.04e+11 &$>$1.40e+10 \\ 
0119$+$115 &Q &0.570 & 18.58 $\pm$   0.82 &$ $3.37e+11 &$ $1.50e+11 &$>$5.90e+10 \\ 
0133$+$476 &Q &0.859 & 15.40 $\pm$   1.20 &$>$1.91e+13 &$ $7.14e+11 &$ $2.40e+11 \\ 
0149$+$218 &Q & 1.32 & 18.40 $\pm$   1.90 &$>$1.40e+12 &$ $5.80e+11 &$ $5.10e+10 \\ 
0202$+$149 &Q &0.405 & 15.88 $\pm$   0.75 &$>$1.53e+12 &$ $1.82e+11 &$>$4.80e+10 \\ 
0202$+$319 &Q &1.466 & 10.10 $\pm$   1.00 &$>$2.67e+12 &$ $1.10e+12 &$ $7.90e+10 \\ 
0212$+$735 &Q &2.367 &  6.58 $\pm$   0.18 &$ $1.08e+12 &$ $2.00e+11 &$>$2.70e+09 \\ 
0224$+$671 &Q &0.523 & 13.69 $\pm$   0.56 &$>$3.94e+11 &$>$3.94e+11 &$>$6.60e+10 \\ 
0234$+$285 &Q &1.207 & 22.00 $\pm$   1.10 &$>$3.83e+12 &$>$4.51e+11 &$ $2.40e+11 \\ 
0238$-$084 &G &0.005 &  0.39 $\pm$   0.02 &$>$3.83e+12 &$>$4.51e+11 &$>$2.20e+10 \\ 
0300$+$470 &B &  ... &  0.00 $\pm$   0.00 &$>$8.46e+11 &$>$8.46e+11 &$ $2.90e+10 \\ 
0316$+$413 &G &0.017 &  0.29 $\pm$   0.02 &$ $1.45e+11 &$ $1.45e+11 &$ $4.70e+10 \\ 
0333$+$321 &Q &1.263 & 13.05 $\pm$   0.16 &$>$5.11e+12 &$ $2.64e+11 &$ $1.40e+11 \\ 
0336$-$019 &Q &0.852 & 24.40 $\pm$   1.60 &$>$3.24e+12 &$ $4.51e+11 &$ $5.60e+10 \\ 
0355$+$508 &Q & 1.52 &  1.90 $\pm$   1.60 &$>$9.77e+11 &$>$6.59e+11 &$ $1.20e+11 \\ 
0415$+$379 &G &0.049 &  8.14 $\pm$   0.31 &$>$7.63e+11 &$ $1.56e+11 &$ $4.10e+10 \\ 
0420$+$022 &Q &2.277 &  8.51 $\pm$   0.98 &$ $5.62e+11 &$ $3.02e+11 &$ $6.70e+10 \\ 
0420$-$014 &Q &0.915 &  5.76 $\pm$   0.59 &$>$5.18e+13 &$>$2.12e+12 &$ $1.90e+11 \\ 
0422$+$004 &B &0.310 &  0.39 $\pm$   0.19 &$>$1.38e+12 &$>$1.36e+12 &$>$5.60e+10 \\ 
0430$+$052 &G &0.033 &  6.43 $\pm$   0.24 &$>$9.19e+11 &$>$1.04e+11 &$ $9.80e+10 \\ 
0440$-$003 &Q &0.844 &  0.59 $\pm$   0.15 &$ $2.31e+11 &$ $6.87e+10 &$>$5.20e+10 \\ 
0458$-$020 &Q &2.291 & 13.56 $\pm$   0.82 &$ $1.82e+12 &$ $8.17e+11 &$ $1.20e+11 \\ 
0528$+$134 &Q & 2.07 & 17.33 $\pm$   0.55 &$>$2.06e+13 &$ $1.11e+12 &$>$2.50e+11 \\ 
0529$+$075 &Q &1.254 & 18.00 $\pm$   1.10 &$>$1.91e+10 &$>$1.48e+10 &$>$1.30e+11 \\ 
0552$+$398 &Q &2.363 &  1.63 $\pm$   0.10 &$ $1.16e+12 &$ $6.01e+11 &$ $2.20e+11 \\ 
0607$-$157 &Q &0.324 &  1.20 $\pm$   0.83 &$>$1.08e+13 &$ $1.49e+12 &$ $2.90e+10 \\ 
0642$+$449 &Q &3.408 &  8.52 $\pm$   0.41 &$>$2.84e+13 &$ $4.32e+12 &$ $1.60e+11 \\ 
0707$+$476 &Q &1.292 & -2.80 $\pm$   1.70 &$ $5.33e+11 &$ $2.67e+11 &$ $7.10e+10 \\ 
0716$+$714 &B &  ... &  0.00 $\pm$   0.00 &$>$1.85e+13 &$>$3.60e+11 &$ $3.60e+11 \\ 
0727$-$115 &Q &1.591 & 31.20 $\pm$   0.60 &$>$9.93e+12 &$>$9.61e+11 &$ $1.40e+10 \\ 
0735$+$178 &B &0.424 &  5.05 $\pm$   0.61 &$>$6.99e+11 &$>$1.34e+11 &$ $2.40e+10 \\ 
0736$+$017 &Q &0.191 & 13.79 $\pm$   0.20 &$>$2.51e+12 &$ $5.09e+11 &$ $1.00e+11 \\ 
0738$+$313 &Q &0.630 & 10.70 $\pm$   1.20 &$>$6.54e+11 &$>$7.37e+10 &$ $3.60e+10 \\ 
0748$+$126 &Q &0.889 & 14.57 $\pm$   0.56 &$>$2.66e+12 &$ $9.06e+11 &$ $1.60e+11 \\ 
0804$+$499 &Q &1.432 &  1.17 $\pm$   0.23 &$>$2.94e+12 &$ $1.04e+12 &$ $6.20e+10 \\ 
0814$+$425 &B &0.530 &  0.00 $\pm$   0.00 &$>$1.40e+12 &$>$1.09e+11 &$ $2.30e+10 \\ 
0823$+$033 &B &0.506 & 12.88 $\pm$   0.49 &$>$3.01e+12 &$>$5.29e+11 &$ $5.00e+10 \\ 
0827$+$243 &Q &0.941 & 19.80 $\pm$   1.30 &$ $1.93e+12 &$ $6.23e+11 &$>$2.10e+11 \\ 
0836$+$710 &Q &2.218 & 21.10 $\pm$   0.77 &$ $6.18e+12 &$ $2.97e+11 &$>$1.70e+11 \\ 
0850$+$581 &Q &1.322 & 12.70 $\pm$   4.10 &$>$1.09e+11 &$>$5.30e+10 &$ $3.20e+10 \\ 
0851$+$202 &B &0.306 & 15.13 $\pm$   0.43 &$>$4.80e+12 &$ $5.08e+11 &$ $2.00e+11 \\ 
0906$+$015 &Q &1.018 & 22.08 $\pm$   0.47 &$ $6.20e+12 &$ $6.43e+11 &$>$1.40e+11 \\ 
0917$+$624 &Q &1.446 & 12.10 $\pm$   1.20 &$ $2.74e+11 &$ $9.44e+10 &$ $4.00e+10 \\ 
0945$+$408 &Q &1.252 & 20.21 $\pm$   0.95 &$>$3.07e+12 &$ $1.67e+11 &$ $5.00e+10 \\ 
0954$+$658 &B &0.367 & 12.74 $\pm$   0.83 &$>$9.06e+11 &$>$9.06e+11 &$>$9.50e+10 \\ 
\enddata
\end{deluxetable}

\setcounter{table}{0}
\begin{deluxetable}{cccrrrr}
\tablecolumns{7}
\tablewidth{0pc}
\tablecaption{(Continued)\label{tbl1-cont}}
\tablehead{
\colhead{} & 
\colhead{} & 
\colhead{} &
\colhead{$\beta_{\rm app}$}    & 
\colhead{$T_{\rm b}^{15~{\rm GHz,max}}$}    & 
\colhead{$T_{\rm b}^{15~{\rm GHz,25\%med}}$}    & 
\colhead{$T_{\rm b}^{86~{\rm GHz}}$}\\
\colhead{Name} & 
\colhead{Type} & 
\colhead{$z$} &
\colhead{[$c$]}    & 
\colhead{[K]}    & 
\colhead{[K]}    & 
\colhead{[K]}
}
\startdata
%here Ba-Tb.2.tableout-p2
1012$+$232 &Q &0.565 &  9.00 $\pm$   0.60 &$>$2.48e+12 &$ $4.13e+11 &$ $3.90e+10 \\ 
1101$+$384 &B &0.031 &  0.28 $\pm$   0.05 &$ $2.19e+11 &$ $6.14e+10 &$ $2.10e+10 \\ 
1128$+$385 &Q &1.733 &  1.10 $\pm$   0.50 &$>$4.94e+12 &$ $5.30e+11 &$ $8.10e+10 \\ 
1150$+$497 &Q &0.334 & 17.50 $\pm$   2.00 &$>$4.94e+12 &$ $5.30e+11 &$ $1.40e+11 \\ 
1156$+$295 &Q &0.729 & 24.60 $\pm$   1.90 &$ $4.95e+12 &$ $2.60e+11 &$ $2.60e+11 \\ 
1219$+$285 &B &0.102 &  9.12 $\pm$   0.79 &$>$1.43e+11 &$>$7.39e+10 &$ $1.60e+10 \\ 
1226$+$023 &Q &0.158 & 14.86 $\pm$   0.17 &$>$5.60e+12 &$>$4.73e+11 &$ $8.50e+10 \\ 
1228$+$126 &G &0.004 &  0.03 $\pm$   0.00 &$>$4.52e+11 &$ $7.68e+10 &$ $1.80e+10 \\ 
1253$-$055 &Q &0.538 & 20.57 $\pm$   0.79 &$>$2.42e+13 &$ $5.53e+12 &$>$8.90e+11 \\ 
1308$+$326 &Q &0.997 & 27.50 $\pm$   1.20 &$>$2.93e+12 &$ $2.26e+11 &$ $1.80e+11 \\ 
1502$+$106 &Q &1.833 & 17.55 $\pm$   0.90 &$>$3.22e+12 &$>$1.32e+12 &$>$2.70e+11 \\ 
1508$-$055 &Q &1.191 &  6.20 $\pm$   1.20 &$>$7.84e+11 &$ $3.76e+11 &$ $4.60e+10 \\ 
1510$-$089 &Q &0.360 & 28.00 $\pm$   0.60 &$>$5.60e+12 &$ $3.20e+11 &$>$7.00e+10 \\ 
1546$+$027 &Q &0.412 & 12.10 $\pm$   1.20 &$>$2.76e+12 &$ $2.83e+11 &$>$2.10e+10 \\ 
1548$+$056 &Q &1.422 & 11.60 $\pm$   1.70 &$>$1.03e+12 &$ $3.13e+11 &$>$1.00e+11 \\ 
1606$+$106 &Q &1.226 & 19.08 $\pm$   0.86 &$>$2.65e+12 &$ $2.26e+11 &$>$1.50e+11 \\ 
1637$+$574 &Q &0.751 & 13.61 $\pm$   0.89 &$>$1.33e+12 &$ $8.54e+11 &$ $3.20e+11 \\ 
1642$+$690 &Q &0.751 & 14.53 $\pm$   0.26 &$>$3.19e+12 &$>$3.18e+11 &$ $1.60e+11 \\ 
1652$+$398 &B &0.033 &  0.87 $\pm$   0.21 &$ $6.81e+10 &$ $5.22e+10 &$ $1.40e+09 \\ 
1655$+$077 &Q &0.621 & 14.80 $\pm$   1.10 &$>$7.37e+11 &$ $1.42e+11 &$ $1.00e+11 \\ 
1739$+$522 &Q &1.379 &  9.08 $\pm$   4.64 &$ $3.77e+12 &$ $1.61e+11 &$ $3.90e+11 \\ 
1741$-$038 &Q &1.057 &  6.64 $\pm$   1.74 &$>$2.03e+13 &$ $1.35e+12 &$ $1.90e+11 \\ 
1749$+$096 &B &0.320 &  7.90 $\pm$   0.75 &$>$1.85e+13 &$>$1.74e+12 &$ $6.10e+11 \\ 
1800$+$440 &Q &0.663 & 15.49 $\pm$   0.32 &$>$9.35e+12 &$ $7.02e+11 &$ $1.70e+11 \\ 
1803$+$784 &B &0.680 & 10.80 $\pm$   1.20 &$ $2.71e+12 &$ $6.00e+11 &$ $8.00e+10 \\ 
1807$+$698 &B &0.050 &  0.09 $\pm$   0.01 &$>$3.92e+11 &$ $1.18e+11 &$>$1.00e+11 \\ 
1823$+$568 &B &0.663 & 26.20 $\pm$   2.60 &$>$5.34e+12 &$>$7.04e+11 &$ $1.40e+11 \\ 
1828$+$487 &Q &0.692 & 13.06 $\pm$   0.14 &$>$2.78e+12 &$ $2.55e+11 &$ $2.60e+10 \\ 
1901$+$319 &Q &0.635 &  0.90 $\pm$   0.70 &$ $7.34e+11 &$>$3.21e+11 &$ $3.20e+10 \\ 
1921$-$293 &Q &0.352 &  4.20 $\pm$   1.30 &$ $2.75e+12 &$ $4.49e+11 &$ $3.90e+10 \\ 
1923$+$210 &U &  ... &  0.00 $\pm$   0.00 &$ $2.75e+12 &$ $4.49e+11 &$ $5.60e+10 \\ 
1928$+$738 &Q &0.303 &  8.16 $\pm$   0.21 &$ $2.37e+12 &$ $3.19e+11 &$ $4.70e+10 \\ 
1957$+$405 &G &0.056 &  0.27 $\pm$   0.04 &$ $2.37e+12 &$ $3.19e+11 &$ $2.80e+10 \\ 
2007$+$777 &B &0.342 &  0.30 $\pm$   0.10 &$>$1.06e+12 &$>$1.81e+11 &$>$3.70e+11 \\ 
2013$+$370 &B &  ... & 12.53 $\pm$   0.34 &$>$1.06e+12 &$>$1.81e+11 &$ $1.80e+11 \\ 
2037$+$511 &Q &1.687 &  3.78 $\pm$   0.53 &$ $8.33e+11 &$ $6.07e+11 &$ $4.20e+11 \\ 
2121$+$053 &Q &1.941 & 11.64 $\pm$   0.74 &$>$1.24e+13 &$ $2.63e+12 &$ $8.70e+10 \\ 
2128$-$123 &Q &0.501 &  6.00 $\pm$   0.62 &$>$3.37e+11 &$>$1.29e+11 &$ $2.30e+09 \\ 
2134$+$004 &Q &1.932 &  5.07 $\pm$   0.32 &$>$1.31e+12 &$ $1.97e+11 &$>$4.60e+10 \\ 
2155$-$152 &Q &0.672 & 18.10 $\pm$   1.80 &$>$8.77e+11 &$>$6.53e+11 &$ $2.00e+10 \\ 
2200$+$420 &B &0.069 &  9.95 $\pm$   0.72 &$ $2.26e+12 &$>$5.81e+11 &$>$5.50e+12 \\ 
2201$+$315 &Q &0.298 &  8.28 $\pm$   0.10 &$>$2.77e+12 &$>$1.06e+11 &$ $5.00e+10 \\ 
2216$-$038 &Q &0.901 &  6.75 $\pm$   0.70 &$ $7.04e+11 &$>$1.05e+11 &$ $5.80e+09 \\ 
2223$-$052 &Q &1.404 & 20.33 $\pm$   0.65 &$>$6.53e+12 &$>$1.42e+12 &$ $7.30e+10 \\ 
2234$+$282 &Q &0.795 &  5.10 $\pm$   2.20 &$>$5.13e+11 &$ $4.20e+10 &$ $4.40e+10 \\ 
2251$+$158 &Q &0.859 & 13.79 $\pm$   0.49 &$>$3.37e+12 &$ $4.20e+11 &$>$1.30e+11 \\ 
2255$-$282 &Q &0.927 &  6.00 $\pm$   0.95 &$>$2.24e+13 &$ $4.93e+12 &$ $2.00e+10 \\ 
2345$-$167 &Q &0.576 & 11.46 $\pm$   0.76 &$>$1.39e+12 &$ $2.80e+11 &$ $1.10e+09 \\ 
\enddata
\end{deluxetable}

\section{Intrinsic brightness temperature at 86~GHz}

\subsection{A global 86~GHz VLBI survey data}

In an attempt to investigate intrinsic brightness temperature for sources
observed at 86~GHz,
we used the observed brightness temperatures of VLBI cores at 86\,GHz 
from a large global 86~GHz VLBI survey of compact radio sources~\citep{lee+08}.
The survey data consist of total intensity images with a typical image
FWHM restoring beam of approximately 40~$\mu$as. This corresponds
to a scale of $< 0.1$ parsecs at typical redshifts z$\cong$1 for 
our sample AGNs.
We used Gaussian fit to the VLBI core component of each jet
to determine a rest-frame core brightness temperature $T_{\rm b}$
for each jet according to 

\begin{equation}
\label{tb}
T_{\rm b} =1.22\times 10^{12} \frac{S_{\rm tot}}{d^2\nu^2}(1+z) \,\,{\rm K},
\end{equation}  
where $S_{\rm tot}$ is the fitted core flux density in Janskys at
$\nu$=86~GHz, 
$d$ is the FWHM dimensions of the fitted circular core components
in milliarcseconds.
In determining the FWHM $d$ of a core component,
the resolution limit of the determination was taken into 
account. So, the minimum resolvable size of a component in 
an image is given by
\begin{equation}
d_{\rm min}=\frac{2^{1+\beta/2}}{\pi}{\left[{\pi ab\ln2\ln{\frac{SNR}{SNR-1}}}\right]}^{1/2}, 
\end{equation}
where {\it a} and {\it b} are the axes of the restoring beam, 
{\it SNR} is the signal-to-noise ratio, and $\beta$ is a 
weighting function, which is 0 for natural weighting or 2 
for uniform weighting. When $d<d_{\rm min}$, the uncertainties 
should be estimated with $d=d_{\rm min}$.
If $d < d_{\rm min}$, 
then the lower limit of $T_{\rm b}$ is obtained with $d=d_{\rm min}$.

\subsection{VLBA data}

Since there are no reliable measurements of the apparent jet speeds at 86~GHz,
we used the apparent jet speeds from the 2cm survey
and the MOJAVE survey~\citep{kel+04,lis+09,lis+13},
thereby assuming that
the apparent jet speeds at 15 and 86~GHz are similar each other.
We selected the fastest proper motions for each source from the MOJAVE
survey~\citep{lis+09,lis+13},
assuming that the speeds are maximum values of individual sources.
For some sources whose proper motions are not available in the MOJAVE survey,
we used the apparent speeds from the 2cm survey~\citep{kel+04},
and we only considered those speeds that are ranked 
as ``excellent'' (E) or ``good'' (G) by their criteria.
We found that apparent jet speeds at 15~GHz are available
for 98 of the sources in the 86~GHz survey,

In order to constrain the characteristic intrinsic brightness temperature
at 15~GHz for our sample, we obtained the maximum observed brightness
temperatures ($T_{\rm b}^{\rm 15GHz,max}$) and the 25\%-median
brightness temperatures ($T_{\rm b}^{\rm 15GHz,25\%med}$)
from \cite{kov+05}.
Table~\ref{tbl1} lists
the observed brightness temperatures at 15 and 86~GHz for the selected
target sources, with the optical class and redshift obtained from~\cite{vv06}. 
The final sample contains 98 sources, consisting of
7 galaxies, 20 BL Lac objects,
70 quasars, and one unidentified source according to the optical class.

\subsection{Constraining intrinsic brightness temperature}

\begin{figure}[!t]
 	\centering 
	\epsfxsize=8cm
	%\epsfbox{BaTb.eps}
	\epsfbox{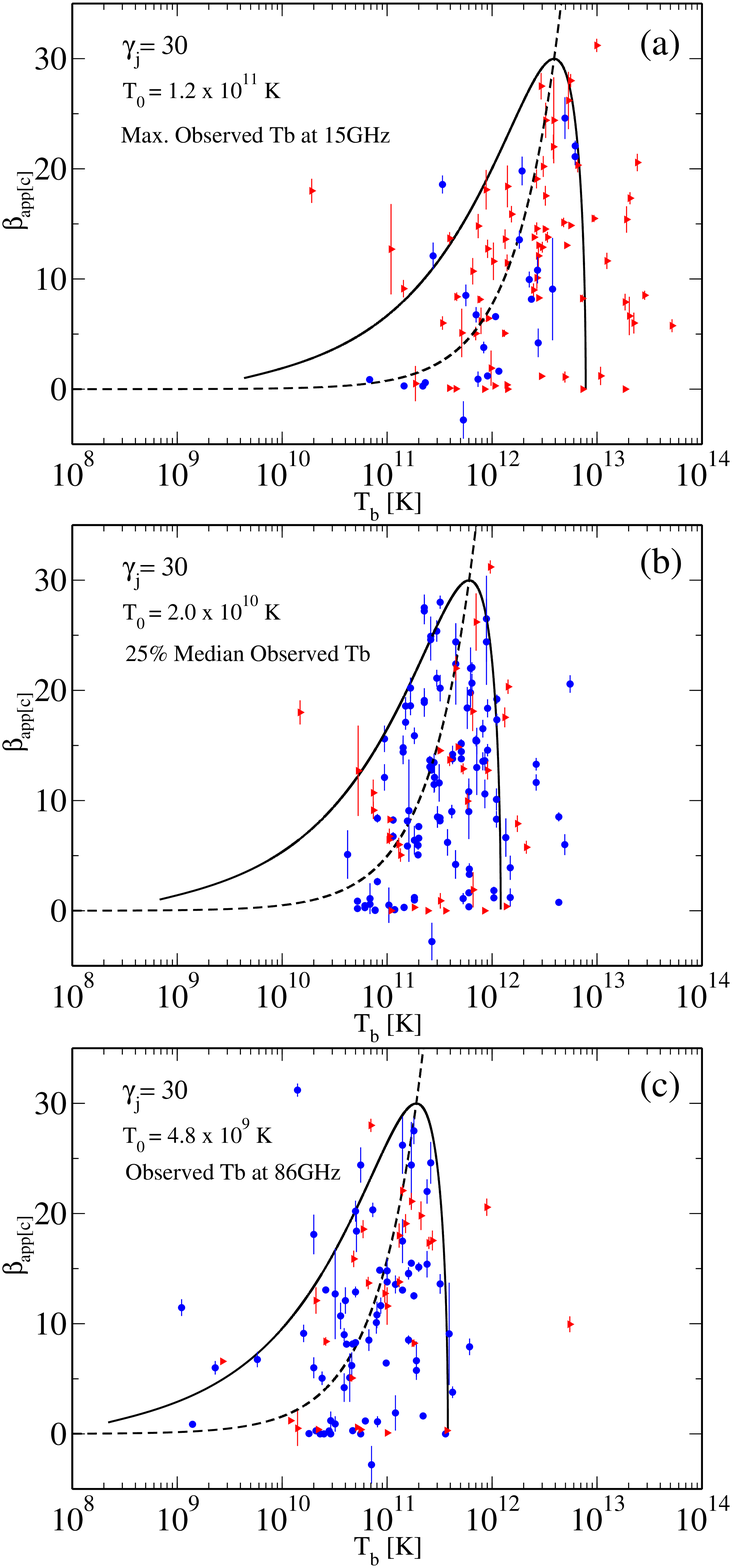}
   	\caption{ 
       Plots of apparent jet speed versus observed brightness
       temperature 
       for the sources in our sample. (a) and (b) plots
       are for the 15~GHz data.
%(a) Maximum observed brightness
%temperatures, and (b) 25\% median observed brightness temperatures.
       (c) plot is for the 86~GHz data. Lower limits of brightness temperatures 
       are indicated by right triangle, and solid circles represent
       measurements.
       The dashed line represents sources observed at the critical angle
       that have the intrinsic brightness temperature of 
       $T_{\rm 0}= 1.2\times 10^{11}$\,K (a),
       $T_{\rm 0}= 2.0\times 10^{10}$\,K (b), and
       $T_{\rm 0}= 4.8\times 10^{9}$\,K (c).
       The solid grey line represents the
       possible apparent speeds of a $\gamma_{\rm j} = 30$ source with
       the corresponding intrinsic brightness temperatures.
   	}
   	\label{fig2} 
\end{figure}

Figure~\ref{fig2} shows plots of the apparent speed 
$\beta_{\rm app}$ versus the observed brightness temperature $T_{\rm b}$ for 
our sample.
The top panel is a plot of $\beta_{\rm app}$ versus the maximum
observed brightness temperatures at 15~GHz for each source in our sample.
The dashed line indicates sources at the critical angle which have 
$T_{\rm 0} = 1.2\times 10^{11}$\,K, and the solid grey line was calculated
using the same value for $T_{\rm 0}$. The value of $T_{\rm 0}$ was chosen
to have
about 75\% of the sources in the right of and below
the dashed line,
which is the same criterion 
for choosing the value of $T_{\rm 0}$ as in~\cite{hom+06}.
The middle panel is
for sources
in their median-low (25\% median) state.
This is the median of the lowest half of the brightness temperature
observations for a given source as defined in \cite{hom+06}.
The 25\% median determined in this critera represents
a typical low brightness state for each source at 15~GHz.
We found a characteristic intrinsic brightness temperature 
of $T_{\rm 0} = 2.0\times 10^{10}$~K
for the 25\% median state. 
The intrinsic temperatures chosen at 15~GHz for our sample (98 sources)
are very close to those for the sample (106 sources) of \cite{hom+06}.
This implies that two samples are statistically similar and hence
suitable for applying this method.

The bottom panel of Figure~\ref{fig2} is a similar plot for the 
observed brightness temperatures at 86~GHz for each source in our sample.
The value of $T_{\rm 0} = 4.8\times 10^{9}$\,K
was chosen to have
about 75\% of the sources in the right of and below
the dashed line.
Since other choices for the simulation parameters described in~\cite{hom+06}
give a very similar distribution of $\beta_{\rm app}$ vs $T_{\rm b}$ and
have fractions between 60\% and 80\% of sources with their viewing angles
smaller than the critical angle,
we take the corresponding values of $T_{\rm 0} = 3.3\times 10^{9}$\,K for 
a 80\% fraction and $T_{\rm 0} = 7.4\times 10^{9}$\,K for a 60\% fraction 
as the range of uncertainty for $T_{\rm 0}$.

\begin{figure}[!t]
 	\centering 
	\epsfxsize=8cm
	\epsfbox{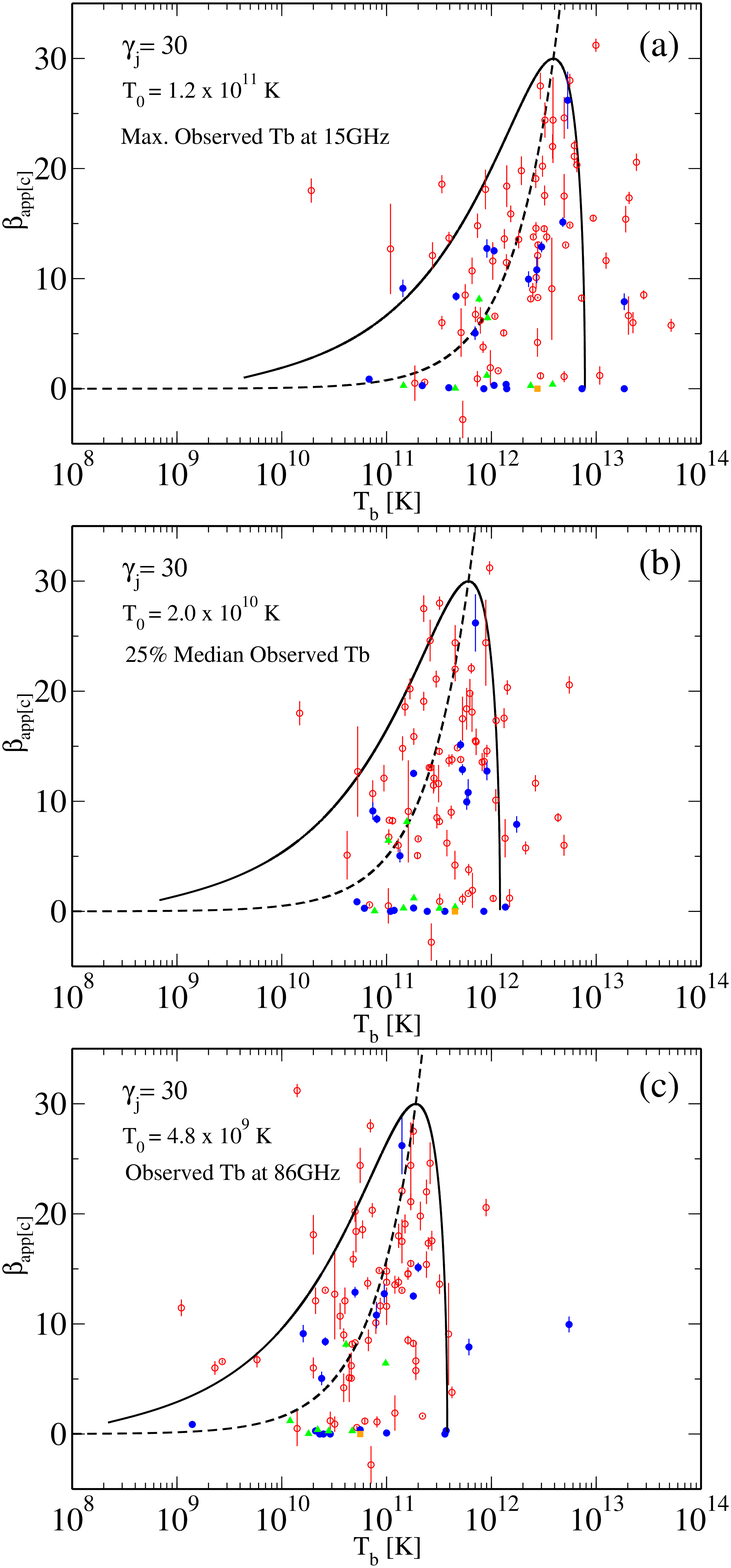}
   	\caption{ 
       Similar plots as Figure~\ref{fig2} except for the symbols:
       red open circles represent quasars;
       blue solid circles represent BL Lac objects;
       green triangles represent galaxies;
       one orange square represents unidentified source.
   	}
   	\label{fig3} 
\end{figure}

Figure~\ref{fig3} shows plots similar to Figure~\ref{fig2} with
sources divided into three groups: quasars, BL Lac objects,
and galaxies. Galaxies have lower apparent jet speeds, whereas
quasars and BL Lac objects are widely spread in speed.

\subsection{Doppler factor}\label{sec:dopplerfactor}

\begin{figure}[!t]
 	\centering 
	\epsfxsize=8cm
	\epsfbox{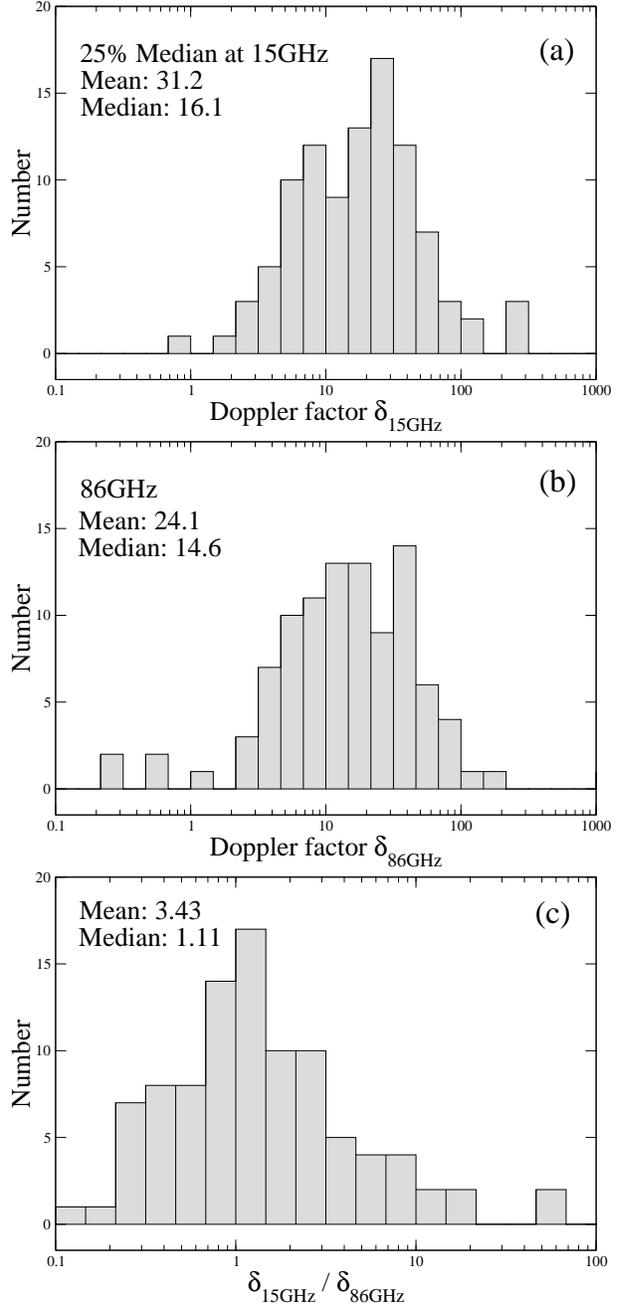}
   	\caption{ 
       Distributions of Doppler factors at (a) 15~GHz and (b) 86~GHz,
       and of (c) their ratio.
       Mean and median values of each distribution are shown in 
       each plot.
   	}
   	\label{fig4} 
\end{figure}

The narrow range of intrinsic brightness temperatures determined at 15~GHz
(median-low state) and 86~GHz enables us to derive Doppler factors
for the sources in our sample according to equation~(\ref{eqn:Tb-3}).
Figure~\ref{fig4}a and Figure~\ref{fig4}b show the distributions of estimated
Doppler factors for the 15~GHz (25\% median) data and the 86~GHz data.
The distributions have the mean values of 31.2 and 24.1
(excluding one value exceeding 1000) with 
the medians of 16.1 and 14.6 for each data, respectively.
This implies that Doppler factors derived with the data at 15~GHz
and 86~GHz for our sample of sources are slightly different from each other.
The Doppler factors of the VLBI core at 86~GHz are lower than
those at 15~GHz.
Figure~\ref{fig4}c shows the distribution of the ratio of Doppler factors
at 15~GHz and 86~GHz for all sources in our sample. The distribution
peaks at a value higher than unity, and has mean and median values of 
3.43 and 1.11, implying that for many sources the estimated Doppler factors
are higher for the 15~GHz jets than for the 86~GHz VLBI cores.
Taking into account equation~(\ref{eqn:delta}),
higher Doppler factors indicate faster apparent jet speeds
for sources whose viewing angles are close to the critical
value $\theta_{\rm c}$ for maximal apparent jet speed.

\section{Discussion}
   
It is interesting that the plots $\beta_{\rm app}$ versus $T_{\rm b}$
in Figures~\ref{fig2}b and ~\ref{fig2}c show similar trends
although few sources are far beyond the solid curve in Figure~\ref{fig2}c. 
There are few sources having low brightness and fast apparent speeds.
As discussed in \cite{coh+07},
this implies that for the fastest jet components in many sources,
the pattern speeds are closely related with the flow speeds.
If the pattern and flow speeds were independent,
then it is difficult to imagine such a trend in $\beta_{\rm app}$ versus
$T_{\rm b}$.
If some sources have fast pattern speeds with slow flow speeds,
then they may be placed to the left of the solid line.
This may be the case for the few sources located in the area
of Figure~\ref{fig2}c. 

As discussed in \cite{hom+06},
the chosen value of $T_{\rm 0}$ should be considered as a lower bound on
the characteristic value. This is because
there are more lower limits of observed brightness temperatures
than measurements in Figure~\ref{fig2}a.  
For our sample 
$T_{\rm 0}>1.2\times10^{11}$~K when sources have their maximum
brightness temperature. Therefore sources should be well away from
equipartition.
However, for the median-low state in Figure~\ref{fig2}b,
there are fewer limits of the observed temperatures than measurements.
We may consider the chosen value of $T_{\rm 0}$ the characteristic
median-low intrinsic brightness temperature. For our sample
$T_{\rm 0}=2.0\times10^{10}$~K. 
This value of $T_{\rm 0}$ is similar to the brightness temperature
under an equipartition condition.

For the 86~GHz data, there are similar number of lower limits as
the median-low plot, so it seems reasonable to take
the characteristic intrinsic brightness temperature
to be $T_{\rm 0} = 4.8\times 10^{9}$\,K. 
This value is lower by a factor of $\sim$4
than the median-low intrinsic brightness temperature at 15~GHz.
We note that the apparent decrease in brightness temperature
between the 15~GHz and 86~GHz cases cannot be attributed
to scatter in the plots, since the real ranges
of the intrinsic brightness temperate in both cases
are $1.6\times 10^{10}~{\rm K} < T_{\rm 0} < 3.1\times 10^{10}$\,K 
and $3.3\times 10^{9}~{\rm K} < T_{\rm 0} < 7.4\times 10^{9}$\,K
for the 15~GHz and 86~GHz cases, respectively, 
using the 60\%- and 80\%-fraction criteria.

This is lower-than-equipartition temperature implying that
the VLBI cores seen at 86~GHz may be representing a jet region
where the magnetic field energy dominates
the total energy in the jet.
Using equation 5 of \citet{rea94},
we estimate that the 86~GHz VLBI core regions have 
$4.5\times10^8$ times more energy in magnetic fields
as in radiating particles.
In these circumstances we may expect
conversion of magnetic field energy into the kinetic energy
of particles in the jet.
Since the 86~GHz VLBI cores should be located the inner regions of the jet
in contrast to the 15~GHz cores and jets, taking into account the opacity
effect of a relativistic jet~\citep{lob07,LZ06},
we may also expect that the intrinsic brightness temperature
will increase as we go down-stream of the jet.
Applying similar estimates to the 15~GHz data,
we found $1/\eta\simeq2.4\times10^3$, implying
the ratio between the energy in the magnetic fields
and in their radiating particles may change by an order of 5
as the relativistic jet moves from the 86~GHz VLBI core regions to 
the 15~GHz jet regions.

The increase of the intrinsic brightness temperature may 
result in the increase of apparent jet speeds from the jet core
of radio galaxies and BL Lac objects as reported by \cite{lis+13}.
Figure 13 of \cite{lis+13} may imply that
the apparent jet speeds at 15~GHz (in the outer region of the jet) 
are faster than those at 86~GHz (in the inner region),
which is consistent with our results of Doppler factor
in Section~\ref{sec:dopplerfactor}.
Although positive correlation of speed with core distance
needs to be confirmed based on a complete AGN sample, 
we may expect that the relativistic jets of AGNs
may accelerate on moving away from the central engine
with corresponding increase in
the intrinsic brightness temperature.
This also agrees with the results of Doppler factors.
Some observational tests for the jet acceleration model
will be discussed in a separate paper.

The difference in the intrinsic temperatures $T_{\rm 0}$ deduced at 15\,GHz
and 86\,GHz may imply that only a small number of sources will be
suitable for VLBI at higher frequencies (e.g., $\geq$ 300\,GHz). 

\section{Conclusions}

\begin{itemize}

	\item The 86~GHz global VLBI survey has yielded the observed
	brightness temperatures for 98 sources with available apparent
	jet speeds and observed brightness temperatures at 15~GHz
	from the 2~cm VLBA survey and MOJAVE program. 
	On applying the $T_0$-constraining method, 
	we find that the intrinsic brightness temperature is
	$T_{\rm 0} = 4.8^{+2.6}_{-1.5}\times 10^{9}$\,K for the VLBI
	cores seen at 86~GHz.
		
	\item The Doppler factors estimated with the constrained
	intrinsic brightness temperatures tend to be higher for 
	the jets seen at 15~GHz than for the VLBI cores seen at 86~GHz.
	It is likely that Doppler factor increases
	down-stream of a relativistic jet. 

	\item The VLBI cores at 86~GHz in our sample may 
	be such that 
	the magnetic field energy in the jet converts into
	the kinetic energy of particles.
	Moving outwards, down stream of the jet, 
	the intrinsic brightness temperature will increase.

\end{itemize}

%--------------------------------------------------------------------
\acknowledgments{
I would like to thank the anonymous referee for important comments
and suggestions which have enormously improved the manuscript.
The VLBA is an instrument of the National Radio Astronomy Observatory,
which is a facility of the National Science Foundation operated under
cooperative agreement by Associated Universities, Inc.
This work was supported by Global Research Collaboration
and Networking program of Korea Research Council of Fundamental Science
\& Technology (KRCF).
}

%--------------------------------------------------------------------

%-------------------------------------------------------------------
\end{document}